\documentclass[10pt]{article}
\usepackage[utf8]{inputenc}
\usepackage[T1]{fontenc}
\usepackage{newtxtext}
\usepackage{geometry}
\usepackage{graphicx}
\usepackage{amsmath,amssymb}
\usepackage{array,tabularx}
\usepackage{microtype}
\usepackage[hidelinks]{hyperref}
\usepackage{float}
\usepackage[section]{placeins}
\usepackage{caption}
\usepackage{fancyhdr}
\usepackage{titlesec}
\usepackage{xcolor}
\titleformat{\section}{\normalfont\bfseries\normalsize}{\thesection}{0.8em}{}
\titleformat{\subsection}{\normalfont\bfseries\normalsize}{\thesubsection}{0.8em}{}
\titlespacing*{\section}{0pt}{11pt}{5pt}
\titlespacing*{\subsection}{0pt}{9pt}{4pt}
\fancypagestyle{plain}{\fancyhf{}\fancyfoot[R]{\thepage}}
\begin{document}
\thispagestyle{empty}
\noindent\includegraphics[width=\linewidth]{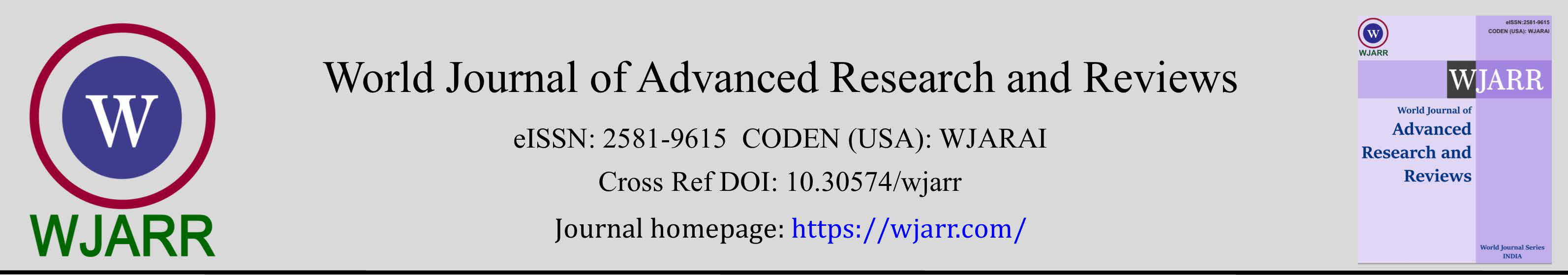}
\vspace{-3pt}
\noindent\begin{minipage}[t]{0.72\linewidth}
{\small\textsc{(Research Article)}}
\end{minipage}\hfill\begin{minipage}[t]{0.26\linewidth}
\raggedleft\includegraphics[width=0.95\linewidth]{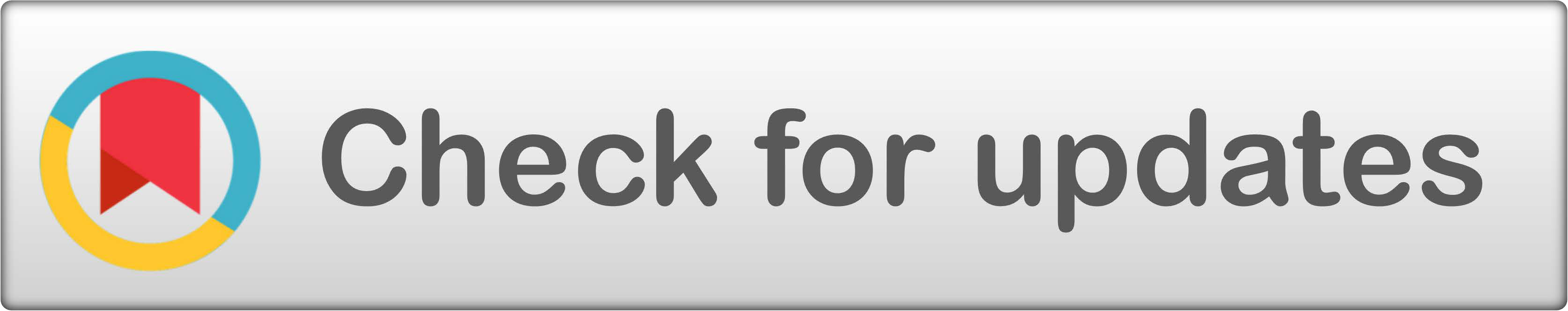}
\end{minipage}
\vspace{4pt}

{\large\noindent Trusted financial data for agentic commerce: A multi-agent framework for credit risk, fraud detection, and secure payment decisions\par}
\vspace{8pt}
{\small Kayode Lateef Ogunsusi\textsuperscript{1}, Mary Nwaife Mezue\textsuperscript{2}, Kazeem Busayo Ogunsusi\textsuperscript{3} and Adeniyi Paul Phillips\textsuperscript{4}\par}
\vspace{8pt}
{\small\itshape
\textsuperscript{1}Credit and Fraud Risk, American Express, Phoenix, Arizona, USA.\par
\textsuperscript{2}Monetization Ads Department, Meta, Washington, USA.\par
\textsuperscript{3}PhD. Department of Statistics, Iowa State University, Iowa, USA.\par
\textsuperscript{4}MS. Computer Science, Austin Peay State University, Tennessee, USA.\par}
\vspace{9pt}
{\small World Journal of Advanced Research and Reviews, 2026, 31(01), 1317--1331\par}
\vspace{5pt}
{\small Publication history: Received on 13 June 2026; revised on 19 July 2026; accepted on 22 July 2026\par}
\vspace{5pt}
{\small Article DOI: \url{https://doi.org/10.30574/wjarr.2026.31.1.1935}\par}
\vspace{7pt}
\hrule\vspace{6pt}
\noindent\textbf{Abstract}\par
\vspace{-5pt}
Financial institutions increasingly require intelligent risk-management systems that combine predictive performance with data reliability, interpretability, and adaptable decision policies for emerging agentic commerce environments. This study proposes a Trusted Financial Data (TFD) Multi-Agent Framework that integrates predictive risk modeling, behavioral intelligence, data-trust assessment, explainable artificial intelligence, and uncertainty-aware decision orchestration for credit-default and payment-fraud management. The framework translates heterogeneous risk and trust signals into interpretable Approve, Review, and Decline decisions using fixed and adaptive decision policies. Experimental evaluation was conducted on the Home Credit Default Risk and IEEE-CIS Fraud Detection datasets against five machine-learning baselines: Logistic Regression, Random Forest, XGBoost, LightGBM, and Multi-Layer Perceptron. Adaptive TFD achieved recall values of 0.6234 and 0.9366, with ROC-AUC values of 0.7584 and 0.9474 on Home Credit and IEEE-CIS, respectively. Adaptive orchestration consistently improved minority-class risk detection relative to fixed TFD policies, although at the cost of increased false-positive interventions. SHAP-based global and local explanations further enhanced transparency into model predictions and decision behavior. Overall, the findings demonstrate the potential of TFD as a trust-aware multi-agent decision-orchestration layer that combines predictive, behavioral, uncertainty, and data-quality signals to support configurable, explainable, and adaptive financial risk management across heterogeneous operational environments.
\par\vspace{6pt}
\par\vspace{6pt}\hrule\vspace{3pt}

\section{Introduction}
The rapid digitalization of financial services has transformed the way consumers, merchants, and financial institutions interact. Advances in Artificial Intelligence (AI), cloud computing, open banking, and real-time payment technologies have enabled financial institutions to automate credit risk assessment, fraud detection, payment authorization, and customer decision support with unprecedented speed and accuracy [1]--[4]. These technologies have improved operational efficiency, enhanced customer experience, and strengthened financial inclusion by enabling intelligent decision-making across increasingly complex payment ecosystems [2], [3].

Despite these advances, the effectiveness of AI-driven financial decision systems remains fundamentally dependent on the quality and trustworthiness of the underlying financial data. Modern financial institutions continuously process high-volume transactional, behavioral, identity, and device information originating from multiple heterogeneous sources. However, fragmented customer records, incomplete transaction histories, inconsistent data governance,

missing or inaccurate records, synthetic identity fraud, and evolving fraud strategies continue to reduce model

reliability and increase operational risk [5]--[9]. These challenges are particularly significant in real-time financial environments where inaccurate decisions may increase fraud losses, delay legitimate customer transactions, and expose institutions to operational, financial, and regulatory risks [6], [8], [10].

The urgency of addressing these challenges is reflected in recent fraud statistics. According to the U.S. Federal Trade Commission (FTC), consumers reported more than US\$12.5 billion in fraud losses during 2024, representing approximately a 25\% increase over the previous year, while investment scams alone accounted for US\$5.7 billion in reported losses [11]. These figures demonstrate that financial fraud continues to evolve in both scale and sophistication, requiring AI systems capable of making reliable, transparent, and timely decisions in increasingly dynamic financial environments [6], [7].

Recent advances in machine learning and deep learning have substantially improved predictive performance for credit risk assessment and fraud detection. Techniques such as Logistic Regression, Random Forest, XGBoost, LightGBM, graph learning, and neural networks have demonstrated strong performance across diverse financial applications [1]--[3], [10], [12]. In parallel, Explainable Artificial Intelligence (XAI) has received increasing attention for improving model transparency and supporting regulatory compliance in high-stakes financial decision-making [13], [14]. Nevertheless, most existing studies primarily focus on improving predictive performance while assuming that the underlying financial data are complete, accurate, and trustworthy. The National Institute of Standards and Technology (NIST) Artificial Intelligence Risk Management Framework (AI RMF 1.0) emphasizes that trustworthy AI depends not only on model performance but also on robust governance, high-quality data, transparency, reliability, and continuous risk management throughout the AI lifecycle [8].

An emerging challenge is the evolution of agentic commerce, where autonomous AI agents increasingly assist users by discovering products, negotiating services, and initiating financial transactions with limited human intervention [15], [16]. Unlike conventional payment environments, agentic commerce requires intelligent systems to reason over heterogeneous financial data, coordinate multiple decision processes, and generate trustworthy outcomes under stringent security, privacy, and latency requirements [4], [8]. Consequently, improving the trustworthiness of financial data has become equally as important as improving predictive algorithms, particularly for autonomous financial decision-making.

Motivated by these challenges, this paper proposes a Trusted Financial Data (TFD) Multi-Agent Framework that integrates data-trust assessment, predictive risk modeling, behavioral intelligence, explainable AI, and adaptive decision orchestration for credit and fraud risk management. The framework is evaluated using the IEEE-CIS Fraud Detection and Home Credit Default Risk datasets and benchmarked against Logistic Regression, Random Forest, XGBoost, LightGBM, and Multi-Layer Perceptron (MLP) models. By placing trusted financial data at the center of AIenabled decision-making, this study examines how predictive and data-quality signals can be translated into interpretable Approve, Review, and Decline decisions for emerging agentic commerce environments.

\section{Literature review}
Recent advances in Artificial Intelligence (AI) have significantly improved financial decision systems by enabling automated credit risk assessment, fraud detection, payment authorization, and regulatory compliance. Machine learning and deep learning models have demonstrated superior predictive capabilities compared with many traditional statistical approaches. Nevertheless, the effectiveness of these techniques remains dependent on the quality of financial data, model transparency, and the ability to adapt to evolving fraud patterns and autonomous financial ecosystems [1]-- [3], [8]. This section reviews existing research in five key areas that motivate the proposed Trusted Financial Data framework.

\subsection{Artificial Intelligence in Financial Decision Systems}
AI has become an integral component of modern financial services, supporting automated decision-making across lending, fraud prevention, transaction monitoring, and payment processing. Ensemble learning methods such as Random Forest, XGBoost, and LightGBM consistently outperform many conventional statistical models because they effectively capture nonlinear relationships within large financial datasets [1]--[3]. More recently, explainable AI techniques have improved model transparency, helping financial institutions satisfy increasing regulatory and governance requirements [14].

Although predictive performance has improved considerably, most existing studies focus primarily on algorithm development while giving comparatively less attention to the trustworthiness of the financial data supporting these models. Consequently, unreliable or inconsistent data continue to limit operational performance in real-world financial systems [8].

\subsection{Artificial Intelligence for Credit and Fraud Risk}
Credit risk assessment and fraud detection remain the two most widely studied financial AI applications. Machine learning algorithms including Logistic Regression, Random Forest, XGBoost, LightGBM, and neural networks have demonstrated strong predictive performance across numerous benchmark datasets [10], [12]. Similarly, graph-based learning has improved the identification of fraud rings and synthetic identity fraud by modeling relationships among customers, merchants, accounts, and devices.

Despite these advances, existing models generally optimize predictive accuracy independently for credit or fraud tasks. Limited research has investigated unified decision frameworks capable of simultaneously integrating credit assessment, fraud intelligence, and payment authorization while considering the reliability of underlying financial data.

\begin{table}[htbp]
\centering
\caption{Comparison of Existing AI Approaches for Financial Risk Management}
\label{tab:1}
\scriptsize
\begin{tabularx}{\linewidth}{|>{\raggedright\arraybackslash}X|>{\raggedright\arraybackslash}X|>{\raggedright\arraybackslash}X|>{\raggedright\arraybackslash}X|}
\hline
\textbf{Technique} & \textbf{Primary application} & \textbf{Strength} & \textbf{Limitation} \\ \hline
Logistic Regression & Credit Risk & Interpretable & Limited nonlinear learning \\ \hline
Random Forest & Credit \& Fraud & Robust performance & Lower interpretability \\ \hline
XGBoost & Credit \& Fraud & High predictive accuracy & Hyperparameter tuning required \\ \hline
LightGBM & Large-scale prediction & Fast and scalable & Sensitive to noisy data \\ \hline
Graph Neural Networks & Fraud Networks & Captures entity relationships & High computational complexity \\ \hline
\end{tabularx}
\end{table}

\subsection{Trusted Financial Data and Data Governance}
Financial AI systems depend on reliable customer, transaction, identity, and behavioral data collected from multiple heterogeneous sources. However, missing values, duplicate records, inconsistent identities, delayed updates, and poor data governance continue to reduce model reliability and decision quality [8], [9]. These challenges become increasingly significant in real-time payment environments, where inaccurate decisions may result in financial losses or unnecessary transaction declines.

Current literature recognizes the importance of data quality but generally treats data preprocessing as an isolated task rather than a continuous component of intelligent financial decision-making. Consequently, relatively few studies explicitly evaluate how trusted financial data influences downstream AI performance.

\begin{table}[htbp]
\centering
\caption{Trusted Financial Data Dimensions}
\label{tab:2}
\scriptsize
\begin{tabularx}{\linewidth}{|>{\raggedright\arraybackslash}X|>{\raggedright\arraybackslash}X|}
\hline
\textbf{Dimension} & \textbf{Importance} \\ \hline
Completeness & Reduces missing information \\ \hline
Accuracy & Improves prediction reliability \\ \hline
Consistency & Eliminates conflicting records \\ \hline
Timeliness & Supports real-time decisions \\ \hline
Integrity & Protects against data manipulation \\ \hline
Provenance & Increases confidence in data origin \\ \hline
\end{tabularx}
\end{table}

\subsection{Explainable AI and Agentic Commerce}
As AI systems increasingly support high-stakes financial decisions, explainability has become essential for regulatory compliance, model transparency, and stakeholder trust. Techniques such as SHAP and LIME provide interpretable

explanations for complex machine learning models, enabling financial institutions to justify automated decisions [13], [14].

At the same time, the emergence of agentic commerce, where autonomous AI agents initiate and manage financial transactions, introduces new challenges related to trustworthy data, coordinated decision-making, and secure payment authorization. Existing research on agentic commerce remains limited, particularly regarding the integration of trusted financial data with collaborative AI architectures capable of supporting autonomous financial ecosystems.

\subsection{Summary of Existing Literature}
Existing studies demonstrate substantial progress in applying AI to credit risk assessment and fraud detection through advanced machine learning, deep learning, and graph-based techniques. However, three important limitations remain.

First, most studies prioritize improving predictive algorithms while giving limited attention to the quality and trustworthiness of the financial data driving AI decisions. Second, credit risk assessment, fraud detection, and payment authorization are typically developed as independent analytical processes rather than coordinated decision systems. Finally, relatively little research addresses the emerging requirements of agentic commerce, where autonomous AI agents require secure, explainable, and trustworthy financial decision-making.

These limitations motivate the development of the proposed Trusted Financial Data (TFD) framework, which integrates trusted financial data assessment, explainable AI, and collaborative multi-agent intelligence within a unified architecture for credit risk, fraud detection, and adaptive financial decision orchestration.

\section{Research gap and motivation}
\subsection{Limitations of Existing AI-Based Financial Decision Systems}
Recent advances in Artificial Intelligence (AI) have significantly improved credit risk assessment, fraud detection, and payment authorization through sophisticated machine learning and deep learning techniques. However, most existing studies primarily focus on maximizing predictive performance while assuming that the underlying financial data are complete, accurate, and reliable. In practice, financial institutions manage heterogeneous data from multiple internal and external sources that frequently contain inconsistencies, missing values, delayed updates, and conflicting customer information. Consequently, even highly accurate predictive models may produce unreliable financial decisions when the underlying data cannot be trusted.

\subsection{Fragmented Financial Decision Processes}
Current financial AI solutions generally address credit risk assessment, fraud detection, identity verification, behavioral analytics, and payment authorization as independent analytical processes. While each component may perform effectively on its own, limited coordination among these systems often results in inconsistent risk assessments, duplicated analyses, and delayed operational decisions. As financial ecosystems become increasingly interconnected, integrated decision intelligence is becoming essential.

\subsection{Emerging Requirements for Agentic Commerce}
The rapid emergence of agentic commerce introduces new demands for autonomous financial decision-making. AI agents are increasingly expected to initiate, evaluate, and authorize financial transactions with minimal human intervention while maintaining security, transparency, and regulatory compliance. Existing research provides limited support for collaborative multi-agent architectures capable of continuously validating financial data, coordinating specialized analytical agents, and producing explainable real-time decisions.

\subsection{Motivation for the Proposed Framework}
These limitations motivate the development of the proposed Trusted Financial Data (TFD) Multi-Agent Framework. Unlike existing approaches that primarily emphasize predictive performance, the proposed framework places trusted financial data at the center of intelligent financial decision-making by integrating data validation, identity intelligence, behavioral analytics, explainable AI, and collaborative multi-agent reasoning within a unified architecture. This approach aims to improve the reliability, transparency, scalability, and adaptability of credit risk assessment, fraud detection, and trust-aware payment decision orchestration for next-generation agentic commerce.

\begin{figure}[!htbp]
\centering
\includegraphics[width=.95\linewidth,height=.67\textheight,keepaspectratio]{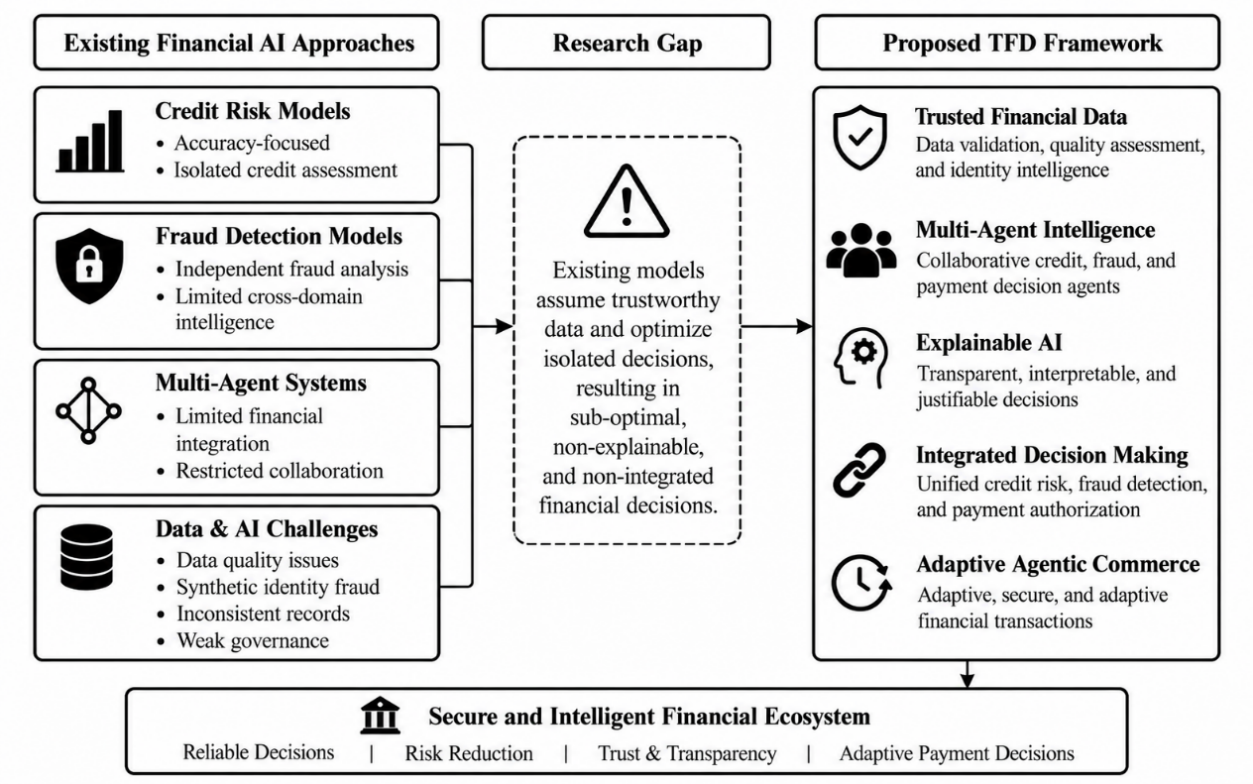}
\caption{Research Gap Addressed by the Proposed Framework}
\label{fig:1}
\end{figure}

\section{Proposed multi-agent framework}
The proposed Trusted Financial Data (TFD) Multi-Agent Framework is designed to improve the reliability, transparency, and adaptability of AI-enabled financial decision systems for emerging agentic commerce. Unlike conventional approaches that independently perform credit risk assessment and fraud detection, the proposed framework integrates trusted financial data, collaborative multi-agent intelligence, and explainable AI into a unified architecture for adaptive financial risk and payment decision orchestration.

\begin{figure}[!htbp]
\centering
\includegraphics[width=.95\linewidth,height=.67\textheight,keepaspectratio]{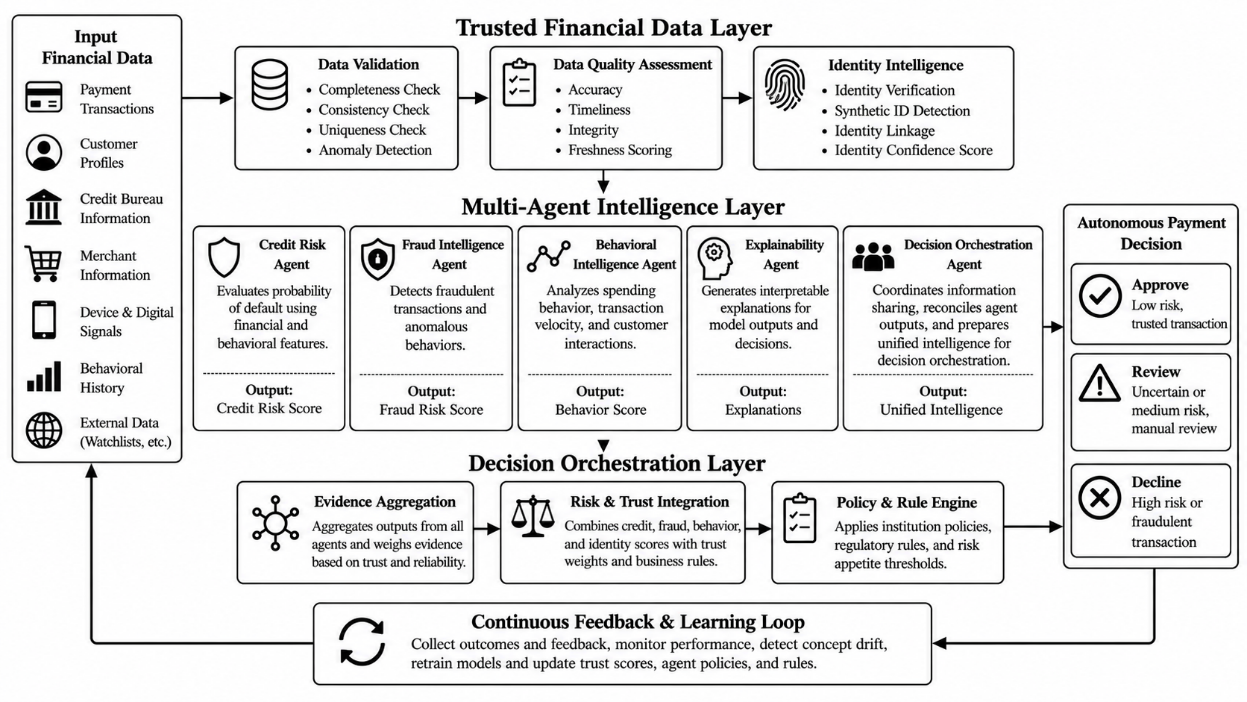}
\caption{Overall architecture of the proposed Trusted Financial Data (TFD) Multi-Agent Framework}
\label{fig:2}
\end{figure}

As illustrated in Figure 2, the framework begins by ingesting heterogeneous financial data from multiple internal and external sources, including payment transactions, customer profiles, credit bureau records, merchant information, device signals, behavioral histories, and external watchlists. Rather than forwarding raw information directly to predictive models, the Trusted Financial Data Layer performs data validation, quality assessment, and identity intelligence to ensure that only reliable information progresses through the analytical pipeline.

The validated financial data are subsequently processed by the Multi-Agent Intelligence Layer, where specialized intelligent agents collaboratively evaluate different aspects of financial risk. The Credit Risk Agent estimates the likelihood of customer default using financial, transactional, and behavioral attributes, while the Fraud Intelligence Agent identifies suspicious activities by detecting anomalous transaction patterns and potential fraudulent behaviors. The Behavioral Intelligence Agent further analyzes customer spending patterns, transaction velocity, and historical interactions to provide additional contextual intelligence that strengthens both credit and fraud assessments. To promote transparency and regulatory compliance, the Explainability Agent generates interpretable explanations for the outputs produced by the analytical agents. These individual assessments are then communicated to the Collaboration and Coordination Agent, which facilitates information sharing, resolves potential conflicts among agent outputs, and establishes a unified representation of financial risk before forwarding the consolidated intelligence to the decision orchestration layer.

The Decision Orchestration Layer integrates the outputs generated by the intelligent agents through evidence aggregation, risk and trust integration, and policy-driven decision analysis. During this process, credit risk, fraud risk, behavioral intelligence, and identity confidence are jointly evaluated alongside institutional policies, regulatory requirements, and predefined risk thresholds to produce a consistent and trustworthy decision. The framework subsequently generates one of three autonomous payment outcomes: Approve, Review, or Decline, depending on the overall risk profile of the transaction. Finally, a Continuous Feedback and Learning Loop captures decision outcomes and operational feedback to support continuous model refinement, trust score recalibration, policy updates, and adaptation to emerging fraud patterns. This feedback mechanism enables the proposed framework to maintain reliable, explainable, and adaptive financial decision-making capabilities in dynamic agentic commerce environments.

To further clarify the responsibilities of the intelligent agents within the proposed framework, Table 3 summarizes the primary function, input, and output of each agent. Although the agents operate independently, they collaboratively exchange validated information to support trustworthy, explainable, and real-time financial decision-making.

\begin{table}[htbp]
\centering
\caption{Functional Description of the Proposed Intelligent Agents}
\label{tab:3}
\scriptsize
\begin{tabularx}{\linewidth}{|>{\raggedright\arraybackslash}X|>{\raggedright\arraybackslash}X|>{\raggedright\arraybackslash}X|>{\raggedright\arraybackslash}X|}
\hline
\textbf{Agent} & \textbf{Primary function} & \textbf{Input} & \textbf{Output} \\ \hline
Credit Risk Agent & Estimates the probability of customer default using financial, transactional, and behavioral information. & Trusted financial data & Credit risk score \\ \hline
Fraud Intelligence Agent & Detects fraudulent transactions through anomaly detection and transaction risk analysis. & Transaction and customer activity data & Fraud risk score \\ \hline
Behavioral Intelligence Agent & Evaluates customer spending behavior, transaction velocity, and historical behavioral patterns. & Behavioral and transactional data & Behavioral risk score \\ \hline
Explainability Agent & Generates interpretable explanations to support transparency, regulatory compliance, and human oversight. & Outputs from analytical agents & Explainability report \\ \hline
Decision Orchestration Agent & Coordinates outputs from all intelligent agents and prepares unified intelligence for policy-based decision orchestration. & Agent outputs & Unified intelligence \\ \hline
\end{tabularx}
\end{table}

Table 3 summarizes the functional roles of the principal intelligent agents within the proposed multi-agent framework. The experimental implementation evaluates a functional subset of these components, focusing on predictive credit or fraud risk estimation, record-level data trust, behavioral-risk integration, fixed and adaptive decision orchestration, and SHAP-based explainability. Identity intelligence, continuous feedback, and broader inter-agent coordination remain architectural extensions for future production deployment.

\section{Framework formulation}
The proposed Trusted Financial Data (TFD) Multi-Agent Framework mathematically formalizes the relationships among trusted financial data, collaborative intelligent agents, and policy-driven decision orchestration for real-time credit risk assessment, fraud detection, and autonomous payment decisions. Unlike conventional financial AI models

that rely primarily on isolated predictive scores, the proposed formulation integrates data trustworthiness, multi-agent intelligence, and explainable decision reasoning into a unified computational framework. This formulation establishes the theoretical foundation for the experimental evaluation presented in the subsequent section. Let the trusted financial dataset be represented as:

\begin{equation}D_T=\{x_1,x_2,\ldots,x_n\}\end{equation}
where DT denotes the validated financial dataset after data quality assessment, identity verification, and integrity validation, while xi represents an individual financial transaction or customer record.

The framework evaluates the trustworthiness of each financial record using a composite trust score that combines multiple data quality dimensions. The trust score for transaction xi is defined as

\begin{equation}T(x_i)=w_1C_i+w_2Q_i+w_3I_i+w_4P_i\end{equation}
where: Ci is the completeness score

Qi represents the data quality score Ii is the identity confidence score Pi denotes the data provenance score

wj is the weighting coefficient associated with each trust dimension.

The weighting coefficients satisfy the normalization constraint

\begin{equation}\sum_{j=1}^{4}w_j=1\end{equation}
ensuring that the composite trust score remains bounded.

Rather than discarding low-trust records, the experimental implementation converts the trust score into a complementary uncertainty signal:

\begin{equation}U(x_i)=1-T(x_i)\end{equation}
where higher values of U(xi) indicate lower confidence in the completeness or reliability of the available data. This uncertainty signal is subsequently incorporated into decision orchestration, enabling lower-trust observations to receive more conservative treatment.

\subsection{Credit Risk Formulation}
Following the trust evaluation process, validated financial records are forwarded to the Credit Risk Agent to estimate the probability that a customer or transaction may result in credit default. The credit risk assessment is formulated as a predictive function that operates on trusted financial data, where each observation is evaluated using historical financial, transactional, and behavioral attributes. The estimated credit risk score is expressed as

\begin{equation}R_C(x_i)=f_C(D_T)\end{equation}
where RC(xi) denotes the predicted credit risk score for financial record, xi, DT represents the trusted financial dataset defined in Equation (1), and fc (.) denotes the credit risk prediction model implemented using supervised machine learning algorithms such as Logistic Regression, Random Forest, XGBoost, LightGBM, or Multi-Layer Perceptron (MLP). The resulting score quantifies the likelihood of default and serves as one of the primary inputs to the decision orchestration process.

\subsection{Fraud Risk Formulation}
Simultaneously, the Fraud Intelligence Agent evaluates each validated financial record to estimate the probability of fraudulent activity. Unlike credit risk assessment, fraud detection focuses on identifying anomalous transaction

behavior, suspicious payment patterns, synthetic identities, and other indicators of financial crime. The fraud risk function is defined as

\begin{equation}R_F(x_i)=f_F(D_T)\end{equation}
where RF(xi) denotes the fraud risk score associated with transaction xi, and fF (.) represents the fraud detection model trained using trusted financial data. This formulation enables the framework to identify potentially fraudulent transactions while minimizing false positives that may negatively affect legitimate customers.

\subsection{Behavioral Intelligence Formulation}
The Behavioral Intelligence Agent continuously evaluates customer spending habits, transaction frequency, merchant interactions, account usage patterns, and historical behavioral trends to identify deviations from expected financial behavior. The behavioral intelligence score is defined as

\begin{equation}R_B(x_i)=f_B(D_T)\end{equation}
where RB(xi) represents the behavioral risk score and fB (.) denotes the behavioral analytics function. This component strengthens the framework by incorporating contextual behavioral information that improves both fraud detection and credit risk assessment.

\subsection{Decision Orchestration Function}
The outputs generated by the individual intelligent agents are subsequently integrated within the Decision Orchestration Agent to produce a unified financial decision score. Rather than relying on a single predictive model, the proposed framework combines credit risk, fraud risk, behavioral intelligence, and trusted financial data through a weighted aggregation process. The integrated decision score is computed as

\begin{equation}R_O(x_i)=\alpha R_C(x_i)+\beta R_F(x_i)+\gamma R_B(x_i)+\delta U(x_i)\end{equation}
\[\alpha+\beta+\gamma+\delta=1\]
where RO(xi) denotes the integrated risk score, RC(xi), RF(xi), and RB(xi) represent credit, fraud, and behavioral risk signals, respectively, while U(x$_i$) = 1 - T(x$_i$) represents data uncertainty. The coefficients determine the relative contribution of each available component according to the application context.

The active components depend on the financial task. Credit-risk signals dominate the Home Credit implementation, whereas fraud-risk signals constitute the primary predictive component in the IEEE-CIS implementation.

\subsection{Adaptive Payment Decision Rule}
The integrated risk score is translated into one of three operational outcomes---Approve, Review, or Decline---using fixed or uncertainty-adjusted thresholds. This threshold-based mechanism enables consistent, explainable, and policydriven financial decision-making while allowing the decision boundaries to adapt to data uncertainty. The adaptive decision rule is defined as

where $\theta$1 and $\theta$2 represent institution-specific decision thresholds that separate low-risk, moderate-risk, and high-risk transactions. This rule enables the proposed framework to generate transparent, consistent, and real-time payment decisions while preserving flexibility for different organizational policies and regulatory requirements.

\begin{equation}D(x_i)=\begin{cases}\mathrm{Approve}, & R_O(x_i)<\theta_1\\\mathrm{Review}, & \theta_1\le R_O(x_i)<\theta_2\\\mathrm{Decline}, & R_O(x_i)\ge\theta_2\end{cases}\end{equation}
where $\theta_1$ and $\theta_2$ represent institution-specific decision thresholds that separate low-risk, moderate-risk, and high-risk transactions. This rule enables the proposed framework to generate transparent, consistent, and real-time payment decisions while preserving flexibility for different organizational policies and regulatory requirements.
\section{Experimental design and case studies}
This section presents the experimental methodology used to evaluate the proposed Trusted Financial Data (TFD) MultiAgent Framework across two complementary financial decision tasks: credit default prediction and payment fraud detection. The evaluation was conducted using the Home Credit Default Risk and IEEE-CIS Fraud Detection benchmark

datasets. These datasets enable assessment of the framework across heterogeneous customer-level credit data and large-scale, highly imbalanced transaction-level fraud data.

The experimental pipeline comprised exploratory data analysis, data preprocessing, baseline model development, TFD framework implementation, adaptive decision orchestration, and explainability analysis. Five supervised machine learning models---Logistic Regression, Random Forest, XGBoost, LightGBM, and Multi-Layer Perceptron (MLP)---were implemented as comparative baselines. Model performance was evaluated using Accuracy, Precision, Recall, F1-score, ROC-AUC, PR-AUC, and Matthews Correlation Coefficient (MCC). The inclusion of PR-AUC and MCC was particularly important because both datasets exhibit class imbalance, for which accuracy alone may provide an incomplete assessment of minority-class predictive performance.

The TFD framework extended predictive risk estimation with data-trust assessment, behavioral intelligence, risk-score integration, and decision orchestration. Two decision configurations were evaluated: a fixed-threshold TFDframework and an adaptive TFD framework in which intervention thresholds respond to record-level data uncertainty. The resulting risk assessments were translated into three operational decision states---Approve, Review, and Decline---to examine the framework's ability to support risk-sensitive financial decision orchestrationbeyond binary classification. Explainability analysis was subsequently performed using SHAP to identify globally influential features and provide local explanations for individual predictions.

\subsection{Financial Datasets}
The experimental evaluation employed two publicly available benchmark datasets representing complementary financial risk domains: The Home Credit Default Risk dataset for credit-risk assessment [17] and the IEEE-CIS Fraud Detection dataset for transaction-level fraud detection [18]. Using both datasets enables the proposed TFD framework to be evaluated across distinct financial decision environments characterized by heterogeneous features and substantial class imbalance.

The Home Credit dataset contained 307,511 observations and 122 original variables, comprising demographic, financial, credit, and application-related attributes used to predict whether an applicant would experience payment difficulties. The dataset represents a highly imbalanced binary classification problem in which default cases constitute the minority class. Following preprocessing, the data were partitioned into stratified training and testing subsets to preserve the original target-class distribution.

The IEEE-CIS dataset combines transaction and identity information using the TransactionID identifier. The merged dataset contained 590,540 transactions and 434 variables. Of these transactions, 20,663 (3.499\%) were labeled as fraudulent, while 569,877 (96.501\%) were legitimate, demonstrating substantial class imbalance. The dataset therefore provides a challenging environment for evaluating fraud-risk detection, data-trust assessment, and adaptive decision orchestration.

Together, the two datasets provide complementary experimental settings: Home Credit evaluates customer-level credit default risk, whereas IEEE-CIS evaluates transaction-level payment fraud risk. This cross-domain design enables assessment of whether the proposed framework can support trust-aware financial decision orchestration across different risk-management applications.

\begin{table}[htbp]
\centering
\caption{Summary of Experimental Datasets}
\label{tab:4}
\scriptsize
\begin{tabularx}{\linewidth}{|>{\raggedright\arraybackslash}X|>{\raggedright\arraybackslash}X|>{\raggedright\arraybackslash}X|>{\raggedright\arraybackslash}X|>{\raggedright\arraybackslash}X|>{\raggedright\arraybackslash}X|}
\hline
\textbf{Dataset} & \textbf{Financial task} & \textbf{Records} & \textbf{Original variables} & \textbf{Minority class} & \textbf{Minority rate} \\ \hline
Home Credit Default Risk & Credit Default Prediction & 307,511 & 122 & Default & 8.07\% \\ \hline
IEEE-CIS Fraud Detection & Transaction Fraud Detection & 590,540 & 434 & Fraud & 3.499\% \\ \hline
\end{tabularx}
\end{table}

variables were separated from predictive features, and stratified train-test splitting was applied using a fixed random state to preserve minority-class representation and ensure consistent experimental partitions.

Numerical and categorical variables were processed separately. Missing numerical values were imputed using statistics estimated from the training data, while categorical missing values were handled before categorical encoding. Numerical features were scaled where required by the learning algorithm, and categorical variables were transformed into machine-readable representations through the fitted preprocessing pipeline. The preprocessing transformations were fitted exclusively on the training partition and subsequently applied to the test partition to reduce the risk of information leakage.

For the IEEE-CIS dataset, transaction and identity records were first merged using TransactionID. Featureauditing was then performed to retain the variables selected for downstream modeling, after which the same deterministic train-test partition was maintained across all baseline and TFD experiments. Because of the substantial class imbalance in both financial datasets, class-sensitive learning strategies were employed rather than relying solely on overall accuracy. Class weighting was used where supported by the respective algorithms, while controlled random undersampling was used for the MLP training workflow to improve computational feasibility and minority-class representation.

No general SMOTE-based oversampling procedure was applied in the final experiments. Model evaluation therefore reflects the preprocessing and imbalance-handling strategies actually implemented in the experimental pipeline. Performance was assessed using Accuracy, Precision, Recall, F1-score, ROC-AUC, PR-AUC, and MCC, with particular emphasis on PR-AUC, Recall, and MCC for evaluating minority-class risk detection.

\subsection{Baseline Models and Experimental Configuration}
Five supervised machine learning models were implemented as comparative baselines: Logistic Regression, Random Forest, XGBoost, LightGBM, and Multi-Layer Perceptron (MLP). These models were selected to provide methodological diversity across linear classification, ensemble learning, gradient boosting, and neural-network-based prediction. All models were evaluated using consistent dataset-specific test partitions to support direct comparison within each experimental case study.

Logistic Regression served as the interpretable linear baseline, while Random Forest provided an ensemble-based approach capable of modeling nonlinear relationships and feature interactions. XGBoost and LightGBM were included as gradient-boosting methods widely applied to structured financial data. The MLP provided a neural-network baseline for learning nonlinear representations from the processed feature space. Class-sensitive configurations were employed where appropriate to address the substantial imbalance between majority and minority classes.

Because the IEEE-CIS dataset was considerably larger and higher-dimensional than the Home Credit dataset, computationally efficient model configurations were used to ensure experimental feasibility. These included parallel processing where supported, histogram-based tree construction for gradient-boosting models, early stopping for the MLP, and controlled model complexity for Random Forest. The MLP workflow additionally used random undersampling of the training data to reduce computational burden while improving minority-class representation. These configurations were applied transparently rather than presenting the experiments as exhaustive hyperparameter optimization.

Each model generated binary class predictions and continuous risk probabilities. Performance was evaluated using Accuracy, Precision, Recall, F1-score, ROC-AUC, PR-AUC, and Matthews Correlation Coefficient (MCC). Given the pronounced class imbalance, particular attention was given to Recall, PR-AUC, and MCC. Training time was recorded during baseline development as a computational-efficiency indicator; however, baseline training times were not directly compared with TFD inference and orchestration times because these measurements represent different computational stages.

For subsequent TFD evaluation, the strongest-performing predictive model was used as the underlying risk-estimation component where appropriate, while the TFD architecture extended conventional probability estimation through datatrust assessment, behavioral risk intelligence, uncertainty-aware risk integration, and decision orchestration. This design enabled the experiments to distinguish between the predictive capability of the underlying machine learning model and the operational decision-support contribution of the proposed framework.

\subsection{Proposed TFD Framework Implementation}
The proposed TFD framework was implemented as a trust-aware decision orchestration layer that integrates modelgenerated risk probabilities with data-quality and behavioral-risk signals. The Data Trust Agent estimated record-level trust from data completeness, while the Behavioral Intelligence Agent combined predictive risk with data uncertainty to support an integrated TFD risk score. For IEEE-CIS, XGBoost served as the underlying fraud-risk model based on its strong overall PR-AUC, F1-score, and MCC performance.

Two decision configurations were evaluated. The Fixed TFD Framework mapped integrated risk scores into Approve, Review, and Decline decisions using predefined thresholds. The Adaptive TFD Framework dynamically adjusted intervention thresholds according to record-level data uncertainty, increasing sensitivity when decisions were supported by lower-trust data. For binary evaluation, Review and Decline were treated as positive risk interventions. Both configurations were evaluated on the same held-out test partitions as the baseline models using Accuracy, Precision, Recall, F1-score, ROC-AUC, PR-AUC, and MCC, while decision distributions were analyzed to assess operational trade-offs.

\subsection{Explainability Analysis}
SHAP was applied to improve the interpretability of the TFD framework and underlying predictive models. Global SHAP analysis identified the features with the greatest influence on risk predictions, while local explanations were generated for representative True Positive, True Negative, False Positive, and False Negative cases. Decision-level explanations were further used to examine the relative contributions of predictive risk, behavioral intelligence, and data-trust signals. This approach provides transparency into both model predictions and the factors influencing TFD-orchestrated financial decisions.

\FloatBarrier
\section{Results and discussion}
This section presents the experimental results of the baseline models and the proposed TFD framework across the Home Credit and IEEE-CIS datasets. Performance is evaluated using classification and imbalance-sensitive metrics, followed by an analysis of fixed and adaptive TFD decision orchestration and model explainability.

\subsection{Predictive Model Performance}
Table 5 summarizes the performance of the baseline models on the IEEE-CIS Fraud Detection dataset, while Figures 3--6 provide complementary model comparisons across the IEEE-CIS and Home Credit datasets. On IEEE-CIS, LightGBM achieved the highest ROC-AUC (0.9486) and Recall (0.8369), whereas XGBoost achieved the highest PR-AUC (0.6962), F1-score (0.4255), and MCC (0.4586) among the baseline models. Random Forest achieved the highest Accuracy (0.9318) and Precision (0.3009), although its lower Recall (0.7167) illustrates the trade-off between precision and minority-class detection. On Home Credit, XGBoost and LightGBM similarly demonstrated strong discriminative performance, achieving ROC-AUC values of 0.7607 and 0.7606, respectively. Overall, the results highlight the importance of evaluating multiple complementary metrics, particularly PR-AUC, Recall, F1-score, and MCC, rather than relying primarily on accuracy when assessing models on highly imbalanced financial-risk datasets.

\begin{table}[htbp]
\centering
\caption{IEEE-CIS Baseline Model Performance}
\label{tab:5}
\scriptsize
\begin{tabularx}{\linewidth}{|>{\raggedright\arraybackslash}X|>{\raggedright\arraybackslash}X|>{\raggedright\arraybackslash}X|>{\raggedright\arraybackslash}X|>{\raggedright\arraybackslash}X|>{\raggedright\arraybackslash}X|>{\raggedright\arraybackslash}X|}
\hline
\textbf{Model} & \textbf{Precision} & \textbf{Recall} & \textbf{F1} & \textbf{ROC-AUC} & \textbf{PR-AUC} & \textbf{MCC} \\ \hline
Logistic Regression & 0.1357 & 0.7295 & 0.2289 & 0.8596 & 0.4136 & 0.2639 \\ \hline
Random Forest & 0.3009 & 0.7167 & 0.4238 & 0.9152 & 0.6071 & 0.4363 \\ \hline
XGBoost & 0.2864 & 0.8268 & 0.4255 & 0.9476 & 0.6962 & 0.4586 \\ \hline
LightGBM & 0.2667 & 0.8369 & 0.4045 & 0.9486 & 0.6940 & 0.4428 \\ \hline
MLP & 0.1696 & 0.7464 & 0.2764 & 0.8851 & 0.4847 & 0.3126 \\ \hline
\end{tabularx}
\end{table}

\begin{figure}[!htbp]
\centering
\includegraphics[width=.89\linewidth,height=.67\textheight,keepaspectratio]{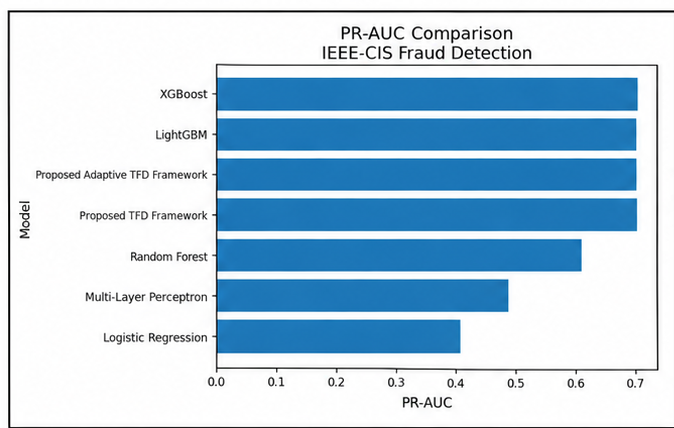}
\caption{PR-AUC comparison of baseline and TFD-based approaches on the IEEE-CIS Fraud Detection dataset}
\label{fig:3}
\end{figure}

\begin{figure}[!htbp]
\centering
\includegraphics[width=.89\linewidth,height=.67\textheight,keepaspectratio]{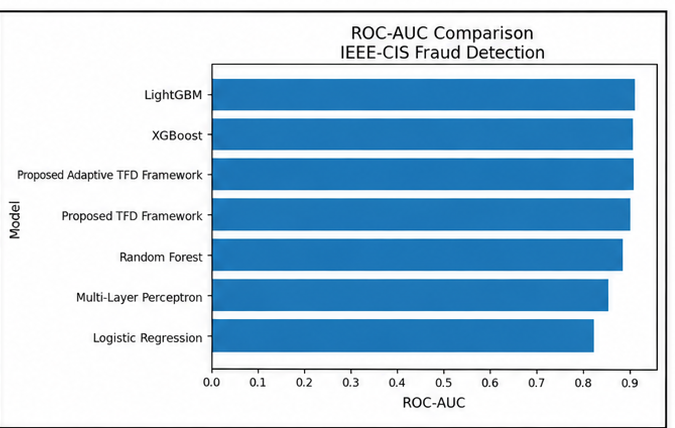}
\caption{ROC-AUC comparison of baseline and TFD-based approaches on the IEEE-CIS Fraud Detection dataset}
\label{fig:4}
\end{figure}

\begin{figure}[!htbp]
\centering
\includegraphics[width=.89\linewidth,height=.67\textheight,keepaspectratio]{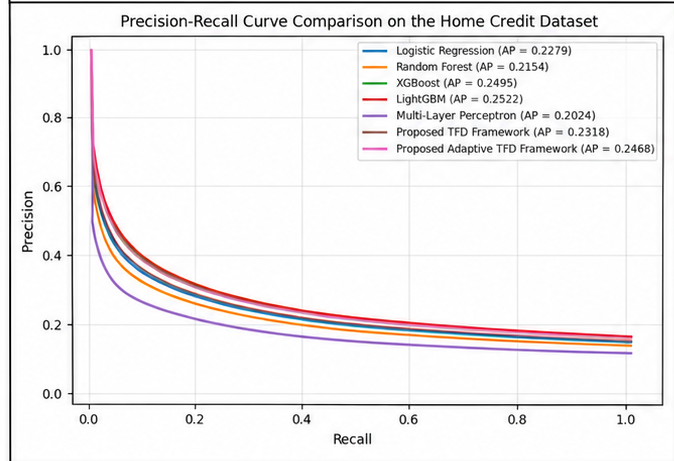}
\caption{Precision-recall curve comparison of baseline models and TFD frameworks on the Home Credit Default Risk dataset}
\label{fig:5}
\end{figure}

\begin{figure}[!htbp]
\centering
\includegraphics[width=.89\linewidth,height=.67\textheight,keepaspectratio]{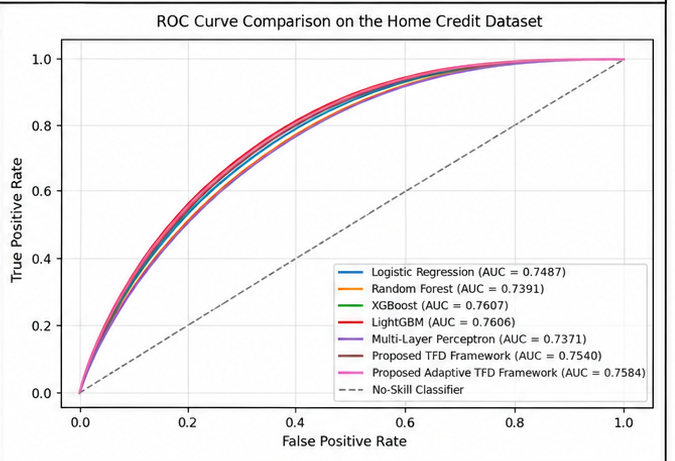}
\caption{ROC curve comparison of baseline models and TFD frameworks on the Home Credit Default Risk dataset}
\label{fig:6}
\end{figure}

\begin{table}[htbp]
\centering
\caption{IEEE-CIS TFD Framework Performance}
\label{tab:6}
\scriptsize
\begin{tabularx}{\linewidth}{|>{\raggedright\arraybackslash}X|>{\raggedright\arraybackslash}X|>{\raggedright\arraybackslash}X|>{\raggedright\arraybackslash}X|>{\raggedright\arraybackslash}X|>{\raggedright\arraybackslash}X|>{\raggedright\arraybackslash}X|}
\hline
\textbf{Framework} & \textbf{Precision} & \textbf{Recall} & \textbf{F1} & \textbf{ROC-AUC} & \textbf{PR-AUC} & \textbf{MCC} \\ \hline
Fixed TFD & 0.1683 & 0.9015 & 0.2837 & 0.9474 & 0.6938 & 0.3485 \\ \hline
Adaptive TFD & 0.1216 & 0.9366 & 0.2153 & 0.9474 & 0.6938 & 0.2863 \\ \hline
\end{tabularx}
\end{table}

\begin{figure}[!htbp]
\centering
\includegraphics[width=.89\linewidth,height=.67\textheight,keepaspectratio]{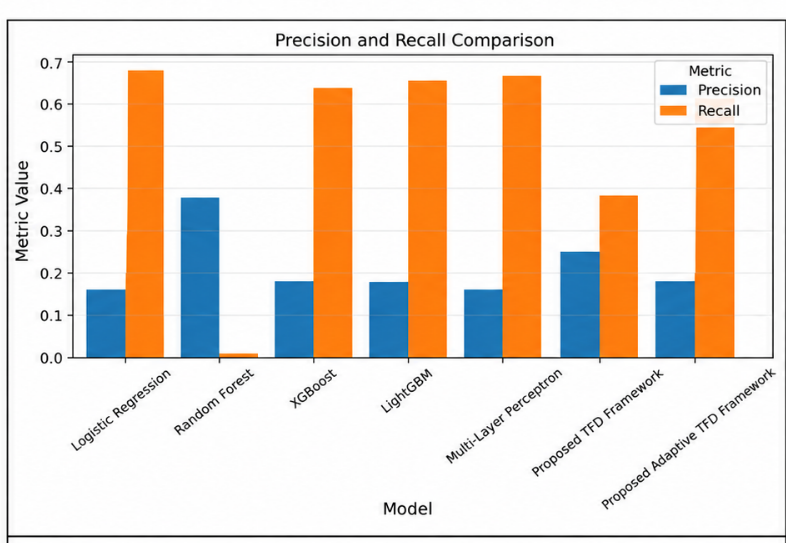}
\caption{Precision and recall comparison of baseline models and TFD frameworks on the Home Credit Default Risk dataset}
\label{fig:7}
\end{figure}

\begin{figure}[!htbp]
\centering
\includegraphics[width=.89\linewidth,height=.67\textheight,keepaspectratio]{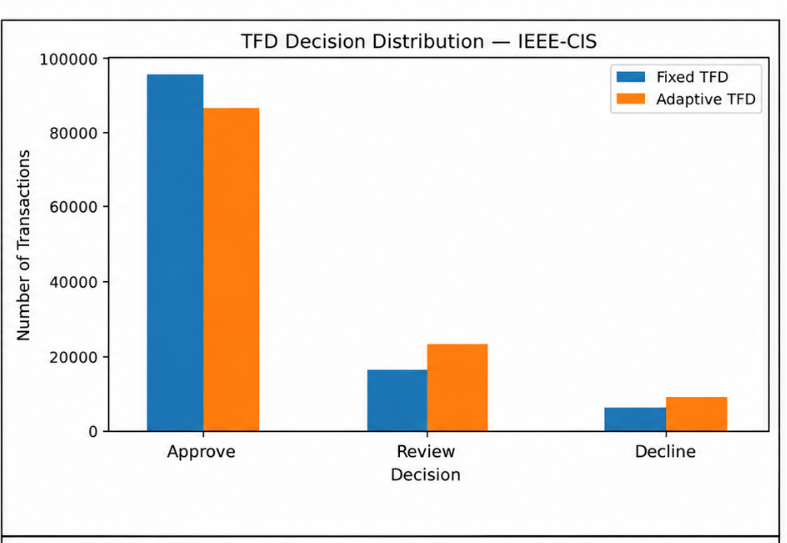}
\caption{Distribution of Approve, Review, and Decline decisions under Fixed and Adaptive TFD orchestration on the IEEE-CIS dataset}
\label{fig:8}
\end{figure}

\FloatBarrier
\subsection{Explainability and Decision Transparency}
SHAP analysis was applied to the Home Credit XGBoost risk model to provide global and local explanations of predictive behavior. Global analysis identified the features exerting the greatest influence on default-risk estimates, while local explanations examined representative True Positive, True Negative, False Positive, and False Negative cases. The TFD decision-level analysis further distinguished the contributions of predictive risk, behavioral intelligence, and data-trust signals, providing transparency into how predictive outputs are translated into operational financial decisions.

\begin{figure}[!htbp]
\centering
\includegraphics[width=.89\linewidth,height=.67\textheight,keepaspectratio]{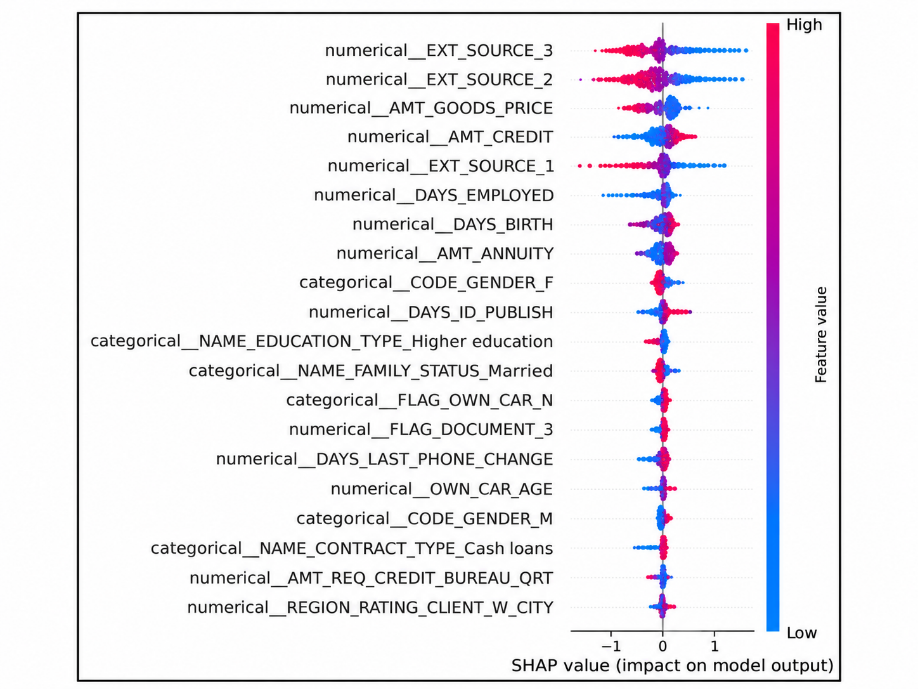}
\caption{Global SHAP summary plot for the Home Credit XGBoost risk model, showing the magnitude and direction of feature contributions to default-risk predictions}
\label{fig:9}
\end{figure}

feature contributions to default-risk predictions

\FloatBarrier
\subsection{Cross-Dataset Discussion and Key Findings}
The cross-dataset evaluation demonstrates that the TFD framework generalizes across complementary financial-risk domains while exhibiting dataset-dependent operating trade-offs. On Home Credit, Adaptive TFD improved default Recall from 0.3837 to 0.6234 relative to Fixed TFD and achieved a ROC-AUC of 0.7584, comparable to XGBoost (0.7607) and LightGBM (0.7606). On IEEE-CIS, Fixed and Adaptive TFD achieved fraud Recall of 0.9015 and 0.9366, respectively, demonstrating strong sensitivity to minority-class risk events.

Across both datasets, adaptive orchestration consistently increased risk detection by applying more conservative interventions under greater data uncertainty, although this improvement was accompanied by reduced precision and

increased false-positive interventions. The findings therefore position TFD primarily as a trust-aware decision orchestration framework rather than a replacement for high-performing predictive models. Its principal contribution is the integration of predictive risk, behavioral intelligence, data trust, and uncertainty-aware decision policies to support interpretable Approve, Review, and Decline actions. This provides financial institutions with configurable operating strategies that can prioritize either balanced intervention or higher risk sensitivity according to application requirements.

\section{Conclusion}
This study proposed a Trusted Financial Data (TFD) Multi-Agent Framework for integrating predictive risk modeling, behavioral intelligence, data trust, and uncertainty-aware decision orchestration in financial risk management. Experimental evaluation on the Home Credit Default Risk and IEEE-CIS Fraud Detection datasets demonstrated the framework's applicability across credit-default and transaction-fraud domains. The results showed that adaptive TFD orchestration consistently increased minority-class recall, achieving Recall of 0.6234 for Home Credit default detection and 0.9366 for IEEE-CIS fraud detection, while maintaining ROC-AUC values of 0.7584 and 0.9474, respectively.

The findings also demonstrate an important operational trade-off: adaptive intervention improves risk sensitivity but increases false-positive interventions, whereas fixed orchestration provides a more balanced decision policy. SHAPbased explainability further supports transparency into influential risk factors and individual predictions. Overall, the TFD framework is positioned not as a replacement for high-performing predictive models, but as a trust-aware orchestration layer that translates predictive and data-quality signals into interpretable Approve, Review, and Decline decisions. Future work will investigate dynamic threshold calibration, temporal validation, concept-drift monitoring, and real-world deployment under evolving financial and regulatory conditions.

\subsection*{Compliance with ethical standards}
\subsection*{Acknowledgments}
The author acknowledges the publicly available Home Credit Default Risk and IEEE-CIS Fraud Detection datasets used for the experimental evaluation in this study.

\subsection*{Disclosure of conflict of interest}
The author declares no conflict of interest.

\subsection*{Statement of ethical approval}
Ethical approval was not required for this study because the research used publicly available benchmark datasets and did not involve direct interaction with human participants.

\subsection*{Statement of informed consent}
Informed consent was not applicable because this study used publicly available benchmark datasets and did not involve direct recruitment or interaction with human participants.

\end{document}